# Strongly Constrained and Appropriately Normed Semilocal Density Functional

Jianwei Sun[1], Adrienn Ruzsinszky[1], and John P. Perdew[1,2]

[1]Dept. of Physics, Temple U., Philadelphia, PA 19122

[2]Dept. of Chemistry, Temple U., Philadelphia, PA 19122

## Abstract

The ground-state energy, electron density, and related properties of ordinary matter can be computed efficiently when the exchange-correlation energy as a functional of the density is approximated semilocally. We propose the first meta-GGA (meta-generalized gradient approximation) that is fully constrained, obeying all 17 known exact constraints that a meta-GGA can. It is also exact or nearly exact for a set of "appropriate norms", including rare-gas atoms and nonbonded interactions. This SCAN (strongly constrained and appropriately normed) meta-GGA achieves remarkable accuracy for systems where the exact exchange-correlation hole is localized near its electron, and especially for lattice constants and weak interactions.

Over the past 50 years, Kohn-Sham density functional theory (KS-DFT) [1-3] has become an *ab initio* pillar of condensed matter physics and related sciences. In this theory, the ground-state electron density $n(\vec{r})$ and total energy $E$ for non-relativistic interacting electrons in a multiplicative external potential can be found exactly by solving selfconsistent one-electron equations, given the uncomputable exact universal exchange-correlation energy $E_{xc}[n]$ as a functional of $n = \sum_{i,\sigma}^{occ} |\psi_{i,\sigma}|^2$, with $\psi_{i,\sigma}$ a KS orbital. This *xc* energy term can be formally expressed as half the Coulomb interaction between every electron and its exchange-correlation hole in a double integral over space [4,5], but in practice its density functional must be approximated. Semilocal functionals approximate it with a single integral and thus are properly size-extensive and computationally efficient, especially for large unit cells, high-throughput materials searches, and *ab initio* molecular dynamics simulations.

Many features of the exact functional $E_{xc}[n]$ are known. Nonempirical functionals, constructed to satisfy exact constraints on this density functional [6-9], are reliable over a wide range of systems (e.g., atoms, molecules, solids, and surfaces), including many that are unlike those for which these functionals have been tested and validated. In this letter, we present a nonempirical semilocal functional that satisfies all known possible exact constraints for the first time, and is appropriately normed on systems for which semilocal functionals can be exact or extremely accurate.

Semilocal approximations can be written as

$$E_{xc}[n_\uparrow, n_\downarrow] = \int d^3r n \varepsilon_{xc}(n_\uparrow, n_\downarrow, \nabla n_\uparrow, \nabla n_\downarrow, \tau_\uparrow, \tau_\downarrow). \quad (1)$$

Here $n_\uparrow(\vec{r})$ and $n_\downarrow(\vec{r})$, the electron spin densities, are the only ingredients of the local spin density approximation (LSDA) [1,10,11-14]. Spin-density gradients are added in a generalized gradient approximation (GGA) [6,14-19], and the positive orbital kinetic energy densities $\tau_\sigma = \sum_i^{occ} \frac{1}{2} |\nabla \psi_{i,\sigma}|^2$ (implicit nonlocal functionals of $n(\vec{r})$)) are further added in a meta-GGA [7-9,20,21]. The broad usefulness of nonempirical semilocal functionals is evidenced by the fact that the PBE GGA construction paper [6] is the 16th most-cited scholarly article of all time [22].

The LSDA was based on what we call an "appropriate norm": It was by construction exact for the only set of electron densities for which it could be exact, the electron gas of uniform spin densities (or those that vary slowly over space). LSDA was surprisingly useful even for solid surfaces and atoms or molecules. But the second-order gradient expansion [14,23], which improves upon LSDA in the slowly-varying limit, was worse than LSDA for real systems, because LSDA satisfies exact constraints that finite-order gradient expansions do not [4,5,6,24]. Non-empirical GGAs like PBE [6] and nonempirical meta-GGAs like TPSS [7] and revTPSS [8] were constructed to achieve higher accuracy by satisfying more exact constraints, and the H atom was added as an appropriate norm for the meta-GGAs. Unlike the GGAs [18], the meta-GGAs need not choose among incompatible constraints.

Despite early successes [25,26,27], the TPSS and revTPSS meta-GGAs were less accurate than the PBE GGA for the critical pressures of structural phase transitions of solids [28,29]. This was due to a spurious order-of-limits problem [30,31], which could be removed [9] if $\tau$ appeared only in the dimensionless variable

$$\alpha = (\tau - \tau_w)/\tau_{unif} > 0 \ , \tag{2}$$

where $\tau^W = |\nabla n|^2/8n$ is the single-orbital limit of $\tau$ and $\tau^{unif} = (3/10)(3\pi^2)^{2/3}n^{5/3}$ is the uniform-density limit. $\alpha$ recognizes covalent single ($\alpha = 0$), metallic ( $\alpha \approx 1$ ) and weak ( $\alpha \gg 1$) bonds [32] (as does the "electron localization function" [33] $1/(1+\alpha^2)$ ). We constructed several interpolations of the exchange energy density [9,34,35] between $\alpha = 0$ and 1, with extrapolation to $\alpha \gg 1$. These abandoned some of the exact constraints satisfied by TPSS and revTPSS. For example, they used a GGA correlation, which is not one-electron self-correlation free. (Note that, in the presence of a paramagnetic current density, meta-GGAs require a gauge correction [36].)

Here we aim to improve the nonempirical meta-GGA by satisfying all known possible exact constraints, including some not satisfied by TPSS and revTPSS. We also add some appropriate norms for which semilocal functionals can be extremely accurate although not exact: rare-gas atoms and nonbonded interactions. Both norms contain information about $0 < \alpha < \infty$, but the latter brings more information about $\alpha \gg 1$. The common feature of all appropriate norms, and a necessary condition for semilocal approximations to be accurate, is that the exact exchange-correlation hole for a considered density remains close to its reference electron. This condition is not satisfied when electrons are shared over stretched bonds, as in stretched $H_2^+$. Fully nonlocal functionals, including global [37] and local [38] hybrids with exact exchange or self-interaction corrections [11,39], often start from a good semilocal functional, and can better describe such bonds at increased computational cost.

There is an expected error cancellation between semilocal exchange and semilocal correlation, since the exact exchange-correlation hole is deeper and more localized near the electron than is the exact exchange hole. Localization of the exact exchange hole for a density is thus a sufficient but not a necessary condition for localization of the exact exchange-correlation hole. In closed-shell atoms and nonbonded interactions, but not in bonded molecules or jellium surfaces, even the SCAN exchange energy is accurate.

The exchange energy for any pair of spin densities is negative, and can be found from that for a spin-unpolarized total density via the exact spin-scaling relation [40]. Thus we only need to construct a meta-GGA for the spin-unpolarized case,

$$E_x[n] = \int d^3 r n \varepsilon_x^{unif}(n) F_x(s,\alpha) \;, \tag{3}$$

where $\varepsilon_x^{unif}(n) = -(3/4\pi)(3\pi^2 n)^{1/3}$ is the exchange energy per particle of a uniform electron gas, $F_x(s,\alpha)$ is the exchange enhancement factor, and

$$s = |\nabla n|/[2(3\pi^2)^{1/3} n^{4/3}] \tag{4}$$

is the dimensionless density gradient. By using these dimensionless variables, we satisfy the correct uniform density-scaling behavior [41].

For $\alpha \approx 1$, we construct an approximate re-summation of the fourth-order gradient expansion (GE4) for exchange [42], valid for slowly-varying densities with small $s$ and $\alpha \approx 1$:

$$F_x^{GE4}(s,\alpha) = 1 + (10/81)s^2 - (1606/18225)s^4 + (511/13500)s^2(1-\alpha) + (5913/405000)(1-\alpha)^2. \tag{5}$$

This PBE-like resummation is

$$h_x^1(s,\alpha) = 1 + k_1 - k_1/(1 + x/k_1), \tag{6}$$

with

$$x = \mu_{AK} s^2 [1 + (b_4 s^2/\mu_{AK}) exp(-|b_4| s^2/\mu_{AK})] + \{b_1 s^2 + b_2(1-\alpha) exp[-b_3(1-\alpha)^2]\}^2. \tag{7}$$

Here $\mu_{AK} = 10/81$, $b_2 = (5913/405000)^{1/2}$, $b_1 = (511/13500)/(2b_2)$, $b_3 = 0.5$, and $b_4 = \mu_{AK}^2/k_1 - 1606/18225 - b_1^2$. For $\alpha = 0$, we impose the strongly-tightened bound $F_x \leq 1.174$ [43], which is satisfied by LDA

($F_x = 1$) but not by PBE, TPSS, or revTPSS: $F_x(s, \alpha = 0) = h_x^0 g_x(s)$ where $h_x^0 = 1.174$ and

$$g_x(s) = 1 - exp[-a_1 s^{-1/2}] \quad . \tag{8}$$

As in the TPSS and revTPSS meta-GGAs, we fit the exact exchange energy of the hydrogen atom, via $a_1 = 4.9479$. To make the exchange energy per particle scale correctly to a negative constant under non-uniform scaling to the true two-dimensional limit [44,45] (as it does not in PBE, TPSS, or revTPSS), we make $F_x$ vanish like $s^{-1/2}$ as $s \to \infty$ [43].

Then we interpolate $F_x$ between $\alpha = 0$ and $\alpha \approx 1$, and extrapolate to $\alpha \to \infty$:

$$F_x(s, \alpha) = \{h_x^1(s, \alpha) + f_x(\alpha)[h_x^0 - h_x^1(s, \alpha)]\} g_x(s), \tag{9}$$

$$f_x(\alpha) = exp[-c_{1\mathrm{x}}\alpha/(1-\alpha)]\theta(1-\alpha) - d_{\mathrm{x}} exp[c_{2\mathrm{x}}/(1-\alpha)]\theta(\alpha-1), \tag{10}$$

and $\theta(x)$ is a step function of $x$. In the spirit of the correction to a different resummed asymptotic series [46], the interpolation/extrapolation gives no correction to our resummed gradient expansion to any power of $\nabla n$ in the slowly-varying limit. There are three parameters ($c_{1\mathrm{x}} = 0.667$, $c_{2\mathrm{x}} = 0.8$, $d_{\mathrm{x}} = 1.24$) in the interpolation /extrapolation, and one ($k_1 = 0.065$) in the resummed gradient expansion, determined by the appropriate norms.

Figure 1 shows the SCAN exchange enhancement factor $F_x$ for a spin-unpolarized density as a function of reduced density gradient $s$ for several values of $\alpha$. Not only does SCAN obey the rigorous bound $F_x \leq 1.174$ for $\alpha = 0$, but it also (and more strongly) obeys the conjectured bound $F_x \leq 1.174$ for *all* $\alpha$ [35,43]. By comparison, the PBE, TPSS, and revTPSS exchange enhancement factors all tend monotonically to the general Lieb-Oxford bound [47] $1.804 = 2.273/2^{1/3}$ as $s \to \infty$ for all $\alpha$. Thus SCAN is radically different from those previous semilocal functionals.

By analogy with $F_x$, we can define an $n$-dependent $F_{xc} = F_x + F_c$, the enhancement over local exchange due to spin polarization, correlation, and semi-locality. The high-density spin-unpolarized limit of $F_{xc}$ is of course $F_x$ of Eq. (3).

The correlation energy is similarly constructed as an interpolation between $\alpha = 0$ and $\alpha \approx 1$, and an extrapolation to $\alpha \to \infty$. The $\alpha \approx 1$ limit uses a PBE-like expression that recovers the second-order gradient expansion for correlation in the slowly-varying limit [14]. The $\alpha = 0$ limit shares the same formula with the $\alpha \approx 1$ limit, with its local part designed just for 1- and 2-electron systems [48]. The $\alpha = 0$ limit makes the correlation energy vanish for any (fully spin-polarized) one-electron density. In the spin-unpolarized case, it satisfies the 2-electron version of the Lieb-Oxford bound [47,48], $F_{xc} \leq 1.67$, and fits the exchange-correlation energy of the He atom. The SCAN correlation energy is by construction non-positive. It properly scales to a finite negative value per electron under uniform density scaling to the high-density limit [44], and to zero like the exchange energy in the low-density limit. Its correlation energy per electron is properly finite (but improperly zero) under non-uniform density scaling to the true two-dimensional limit [44, 45]. The interpolation has three parameters, to be determined by the appropriate norms. All detailed formulas, and a list of all 17 exact constraints plus our appropriate norms, are given in the supplementary material [49]. An important practical feature of our exchange-correlation enhancement factor $F_{xc}$ is that, as functions of $s$, curves for different $\alpha$ do not cross one another strongly (e.g., Fig. 1). In our experience, this condition is needed to achieve selfconsistent solutions by the approach of Neumann, Nobes, and Handy [56].

By recovering GE4, plus the second-order gradient expansion for correlation, we also recover a nearly-exact linear response for a uniform density [57]. Finally, we are able to satisfy the rigorous general Lieb-Oxford bound $F_{xc} \leq 2.215$, as tightened by Chan and Handy [58]. This bound is approached only in the low-density limit, where our $F_{xc}$ properly shows a weak dependence [7,12] on relative spin polarization.

Now there are seven parameters ($c_{1x}$, $c_{2x}$, $d_x$, $k_1$, $c_{1c}$, $c_{2c}$, $d_c$) which are determined by fitting to (1) the large-Z asymptotic coefficients [17, 59] for the exchange energies of rare-gas atoms [15] of atomic number Z,

$$\lim_{Z\to\infty} E_x(Z) = E_x^{LDA} + \gamma_{x1} Z + \gamma_{x2} Z^{2/3}, \qquad (11)$$

(2) the large-Z asymptotic coefficient of the correlation energy of rare-gas atoms [60],

$$\lim_{Z\to\infty} E_c(Z) = E_c^{LDA} + \gamma_{c1} Z, \qquad (12)$$

identified as a key exact constraint for functional approximation [60], (3) the binding energy curve of compressed $Ar_2$ [61] (with a mean absolute error less than 1 kcal/mol for R=1.6, 1.8. and 2.0 Å, bond lengths much smaller than the equilibrium bond length 3.76 Å), as a paradigm of nonbonded interaction (with Kr, another rare-gas atom, as the united-atom limit), and (4) the jellium surface exchange-correlation energy [18,62] at bulk density parameters $r_s$ = 2, 3, 4, and 6 Bohr, within the "range of the possible" set by two recent Quantum Monte Carlo calculations [63,64] and a kernel-corrected random phase approximation calculation [64]. Note that the exact exchange and correlation holes in the jellium surface have long-range parts which cancel one another perfectly [65, 66]. (In Eqs. (11) and (12), we have found the reference coefficients $\gamma_{x1} = -0.2259$ , $\gamma_{x2} = 0.2551$ , $\gamma_{c1} = 0.0388$ by extrapolating accurate energies for Ne, Ar, Kr, and Xe.)

Our calculations to construct and test the SCAN meta-GGA are described next: For the rare-gas atoms, we use accurate Hartree-Fock orbitals [67]. For jellium surfaces, LDA orbitals are used. Our other calculations are selfconsistent. For the $Ar_2$ binding energy curve, we use the Gaussian code [68] with triple-, quadruple- and quintuple-zeta basis sets, extrapolated to the complete basis-set limit. For other molecules, we use the 6-311++G (3df,3pd) basis set. For weak interactions in the S22 set [69], we use the counterpoise correction to reduce the basis-set superposition error. For solids, we use the VASP code [70] with converged plane-wave basis sets and k-space meshes.

Table 1 shows the relative errors of SCAN for $E_x$, $E_c$, and $E_{xc}$ for the rare-gas atoms, in comparison to accurate reference values [15,35,71,72]. The errors in $E_x$ are less than 0.5%, but error cancellation with the much smaller $E_c$ leads to errors in $E_{xc}$ less than 0.1%. This confirms that rare-gas atoms are an appropriate norm. The relative errors of $E_x$ for compressed $Ar_2$ are 0.26%, about the same as for a single Ar atom.

Table 2 shows the error statistics of SCAN and other semilocal functionals for molecules and solids.

For the G3 set [73] of 223 molecules, including some large organic ones, the error is by construction almost minus the error of the atomization energy. For this set, SCAN is much more accurate than the GGAs PBE and especially PBEsol [18], and about as accurate as the meta-GGAs TPSS [7] and M06L [20]. However, M06L has 35 empirical parameters fitted to atomization energies and other chemical data. TPSS has no such empirical parameter, but its complicated form was developed when atomization energies were a gold standard, and may have been indirectly biased by that. (The form of TPSS was complicated by its use of a second dimensionless ingredient built from $\tau$, $z = \tau_w/\tau > 0$.)

Atomization energies of molecules and cohesive energies of solids may not be the most appropriate or important tests of semilocal functionals. There is little statistical correlation [74] between the error that a functional makes for atomization energies and its error for reaction energies. (1) Most atoms that bind into molecules or solids are open-shell and at least partly spin-polarized, while most molecules and solids are spin-unpolarized. (2) Most chemical reaction energies and all heats of formation from the standard states of the elements, when calculated *ab initio*, do not involve free atoms. Thus spin-polarization errors are more troublesome for atomization energies than for reaction energies. It is most important that the functionals should predict energy differences among molecules and solids at fixed atomic composition [75, 76], e.g., $2H_2O \rightarrow 2H_2 + O_2$. We have verified that SCAN is much better than TPSS or PBE for the energy differences between the diamond and beta-tin structures of solid Si under pressure, and we will test SCAN for other structural phase transformations and for the heats of formation of molecules and solids in future work.

To see that SCAN may give a more consistent description of molecular energies than other semilocal functionals, we define the G3$^{HC}$ set of 46 hydrocarbon molecules. For each tested functional, we subtract from the energy of the partly spin-polarized C atom the average over G3$^{HC}$ of the functional's error per C atom. After this correction, the MAE is much smaller for SCAN than for any other tested functional.

The BH76 set [77] comprises 76 barrier heights for chemical reactions (of order 0 to 50 kcal/mol). The barrier arises at a transition state with long, weak bonds, and full nonlocality can improve it substantially. Nevertheless, SCAN gives better barrier heights than any functional in Table 2 except the meta-GGA M06L, which was partly fitted to barrier heights.

S22 [69] is a set of 22 weak interaction energies (hydrogen and van der Waals bonds, with equilibrium binding energies from about 0 to 20 kcal/mol) between closed-shell complexes. For these energies, SCAN is much better than other functionals (and competes with M06L, which was fitted in part to weak interactions). We believe that this success is related to our appropriate norming. (Of course, no semilocal functional can capture the long-range part of the van der Waals interaction, but SCAN captures much of the intermediate-range part, as M06L does.)

LC20 [78] is a set of 20 lattice constants of solids (from 3.451 to 6.042 Å). For this set, SCAN is far more accurate than any other functional in Table 2. Far less accurate is M06L, which was fitted to molecular data. We expected SCAN to be accurate for lattice constants: Fuchs and Scheffler [79] established that lattice-constant errors arise from the region of core-valence overlap [9].

In summary, we have constructed the first meta-GGA that satisfies all known possible exact constraints (about 6 for exchange, 6 for correlation, and 5 for the sum of the two [49]). But there are still infinitely many ways to satisfy these constraints. Thus we have also satisfied appropriate norms, for which our SCAN meta-GGA can be extremely accurate: the energies of rare-gas atoms and nonbonded interactions. We have not fitted to any real bonded system. Thus we regard our functional as a nonempirical one that can be reliably applied to a wide range of problems unlike those to which it was normed.

Table 2 suggests that SCAN is a major improvement over PBE (and much more so over LSDA), at nearly the same computational cost. In future work, we will further explore the possibilities and limitations of SCAN, which we suspect are close to those of the semilocal form, Eq. (1).

Acknowledgments: This work was supported by NSF under grant DMR-1305135 (JP, JS, and AR). We thank the Center for Computational Science (Tulane) for computer time.

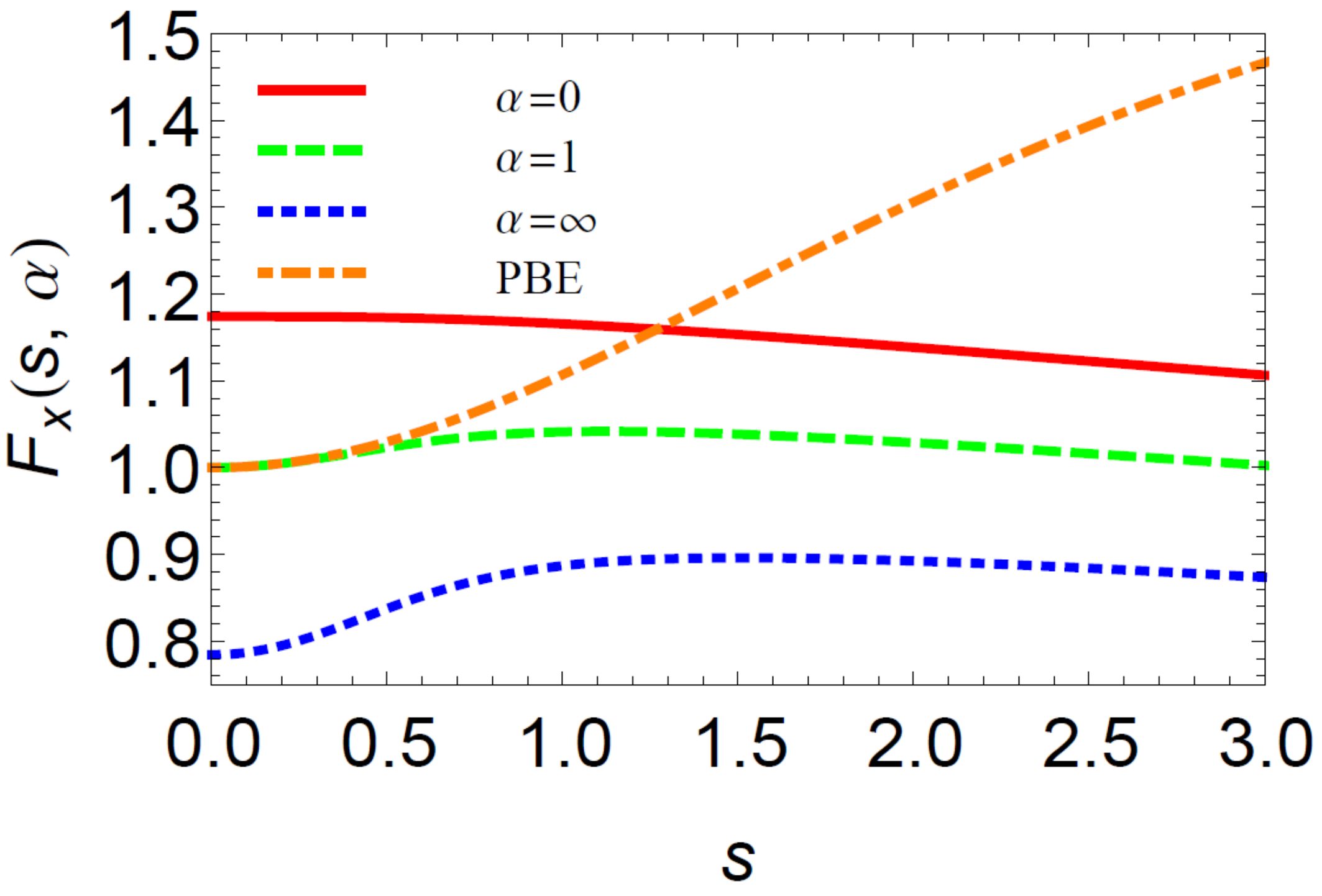

Fig. 1. The SCAN exchange enhancement factor of Eq. (3) for a spin-unpolarized system, as a function of s (the dimensionless density gradient) for several values of α (the dimensionless deviation from a single orbital shape).

Table 1. Relative errors (%) of SCAN for the exchange, correlation, and exchange-correlation energies of the rare-gas atoms.

| | Ne | Ar | Kr | Xe |
|---|---|---|---|---|
| $E_x$ | 0.46 | 0.25 | 0.19 | 0.07 |
| $E_c$ | -11.80 | -4.49 | -5.07 | -3.36 |
| $E_{xc}$ | 0.07 | 0.14 | 0.09 | 0.01 |

Table 2. Mean error (ME) and mean absolute error (MAE) of SCAN and other semilocal functionals for the G3 set of molecules [73], the BH76 set of chemical barrier heights [77], the S22 set of weakly-bonded complexes [69], and the LC20 set of solid lattice constants [78]. For the G3-1 subset of small molecules, the SCAN MAE is 3.2 kcal/mol. $G3^{HC}$ is a subset of 46 G3 hydrocarbons, to which we have applied empirical corrections for the C atom as described in the text to show how consistently SCAN describes molecules. For all data sets, zero-point vibration effects have been removed from the reference experimental values. The LSDA results for G3 are from Ref. [25]. BLYP [15,80], PBEsol [18], and PBE [6] are GGAs; SCAN, TPSS [7], and M06L [20] are meta-GGAs. We could not locate BLYP in VASP, but Ref. [81] suggests that its LC20 MAE may be more than twice that of PBE. (1 kcal/mol = 0.0434 eV)

| | **$G3^{HC}$ (kcal/mol)** | | **G3 (kcal/mol)** | | **BH76 (kcal/mol)** | | **S22 (kcal/mol)** | | **LC20 (Å)** | |
|---|---|---|---|---|---|---|---|---|---|---|
| | ME | MAE | ME | MAE | ME | MAE | ME | MAE | ME | MAE |
| LSDA | -5.6 | 13.0 | -83.7 | 83.7 | -15.2 | 15.4 | 2.3 | 2.3 | -0.081 | 0.081 |
| BLYP | 1.8 | 6.2 | 3.8 | 9.5 | -7.9 | 7.9 | -8.7 | 8.8 | | |
| PBEsol | -4.1 | 6.5 | -58.7 | 58.8 | -11.5 | 11.5 | -1.3 | 1.8 | -0.012 | 0.036 |
| PBE | -2.1 | 6.6 | -21.7 | 22.2 | -9.1 | 9.2 | -2.8 | 2.8 | 0.051 | 0.059 |
| TPSS | 1.9 | 3.8 | -5.2 | 5.8 | -8.6 | 8.7 | -3.7 | 3.7 | 0.035 | 0.043 |
| M06L | -0.2 | 4.6 | -1.6 | 5.2 | -3.9 | 4.1 | -0.9 | 0.9 | 0.015 | 0.069 |
| SCAN | -0.8 | 2.7 | -4.6 | 5.7 | -7.7 | 7.7 | -0.7 | 0.9 | 0.007 | 0.016 |

# Supplementary Material for "Strongly Constrained and Appropriately Normed Semilocal Density Functional"

Jianwei Sun[1], Adrienn Ruzsinszky[1], and John P. Perdew[1,2]

[1]Dept. of Physics, Temple U., Philadelphia, PA 19122

[2]Dept. of Chemistry, Temple U., Philadelphia, PA 19122

We begin by listing all 17 exact constraints, in roughly increasing order of novelty: For exchange, (1) negativity, (2) spin-scaling, (3) uniform density scaling, (4) fourth-order gradient expansion, (5) non-uniform density scaling, and (6) tight bound for two-electron densities. For correlation, (7) nonpositivity, (8) second-order gradient expansion, (9) uniform density scaling to the high-density limit, (10) uniform density scaling to the low-density limit, (11) zero correlation energy for any one-electron spin-polarized density, and (12) non-uniform density scaling. For both together, (13) size extensivity, (14) general Lieb-Oxford bound, (15) weak dependence upon relative spin polarization in the low-density limit, (16) static linear response of the uniform electron gas, and (17) Lieb-Oxford bound for two-electron densities. Constraints (5), (6), (12), and (17) were not satisfied by TPSS [7]. Next we list the appropriate norms: (a) uniform and slowly-varying densities, (b) the jellium surface energy, (c) the H atom, (d) the He atom and the limit of large atomic number for the rare-gas atoms, plus compressed $Ar_2$, and (e) the $Z\rightarrow\infty$ limit of the two-electron ion. Norms (b), (d) and (e) were not used in TPSS.

These supplementary materials also present formulas, figures, and tables that could not fit within the length limits of the main text. (A fuller explication is deferred to future work.) In particular, Table SVI suggests that SCAN might be better than the other tested functionals for the energy differences among hydrocarbon molecules. The hydrocarbons, with transferable bonds which can be described quantitatively by simple empirical rules, seem especially well-suited to the semilocal level of density-functional description.

In Table SIX for the lattice constants of the LC20 solid set, we also include results from three fully-nonlocal vdW-DF functionals [50] designed for solids. This table shows that these vdW-DF's provide reasonably accurate predictions for the lattice constants of solids, while SCAN is significantly better for this property (for which the long-range vdW is not dominant). Of course, vdW-DF's should perform better than SCAN where the long-range vdW interaction is significant. For example, the optB86b vdW-DF only has 0.3 kcal/mol mean absolute error (MAE) for the S22 set, while SCAN has 0.9 kcal/mol MAE, which however is already remarkably good. We expect that SCAN can be improved for both the S22 and LC20 sets by incorporating an appropriate long-range vdW correction.

Now we present the expressions for $\varepsilon_c$, the correlation energy per electron, in detail. We also plot the exchange-correlation enhancement factor $F_{xc}(r_s, \zeta, s, \alpha)$ in the low-density limit ($r_s \to \infty$) and the interpolation/extrapolation functions $f_x(\alpha)$ and $f_c(\alpha)$. Here, $\zeta = (n_\uparrow - n_\downarrow)/(n_\uparrow + n_\downarrow)$ is the spin polarization, $r_s = (4\pi n/3)^{-1/3}$, and $s = |\nabla n|/[2(3\pi^2)^{1/3} n^{4/3}]$ . $\alpha = (\tau - \tau_W)/\tau_{unif}$ with $\tau = \sum_{i,\sigma}^{occ} |\nabla \psi_{i,\sigma}|^2/2$ , $\tau_W = |\nabla n|^2/8n$ , $\tau_{unif} = (3/10)(3\pi^2)^{2/3} n^{5/3} d_s(\zeta)$ , and $d_s(\zeta) = \left[(1+\zeta)^{\frac{5}{3}} + (1-\zeta)^{\frac{5}{3}}\right]/2$. The $\psi_{i,\sigma}$ are Kohn-Sham orbitals.

The semilocal exchange-correlation functional can be written (neglecting $\nabla\zeta$ and assuming that $\alpha$ is the same for spin-unpolarized densities $2n_\uparrow$ and $2n_\downarrow$) as:

$$E_{xc}[n_\uparrow, n_\downarrow] = \int d^3r n\varepsilon_{xc} = \int d^3r n\varepsilon_x^{unif}(n) F_{xc}(r_s, \zeta, s, \alpha), \quad \text{(S1)}$$

where $\varepsilon_x^{unif}(n) = -(3/4\pi)(3\pi^2 n)^{1/3}$ is the exchange energy per electron of a uniform electron gas. The exchange part for a spin-unpolarized density has been given in the main text. The correlation part is:

$$E_c[n_\uparrow, n_\downarrow] = \int d^3r n\varepsilon_c(r_s, \zeta, s, \alpha). \quad \text{(S2)}$$

The SCAN $\varepsilon_c$ has the form:

$$\varepsilon_c = \varepsilon_c^1 + f_c(\alpha)(\varepsilon_c^0 - \varepsilon_c^1), \quad \text{(S3)}$$

where

$f_c(\alpha) = exp[-c_{1c}\alpha/(1-\alpha)]\theta(1-\alpha) - d_c exp[c_{2c}/(1-\alpha)]\theta(\alpha-1),$ (S4)

and $\theta(x)$ is a step function of $x$.

We revise the PBE [6] form for a less-incorrect approach to the two-dimensional limit under nonuniform scaling:

$$\varepsilon_c^1 = \varepsilon_c^{LSDA1} + H_1, \quad \text{(S5)}$$

where

$$H_1 = \gamma\phi^3 \ln[1 + w_1(1 - g(At^2))] \quad \text{(S6)}$$

and $t=(3\pi^2/16)^{1/3}s/(\phi r_s^{1/2})$. Also

$$w_1 = exp[-\varepsilon_c^{LSDA1}/(\gamma\phi^3)] - 1, \quad \text{(S7)}$$

$$A = \beta(r_s)/(\gamma w_1), \quad \text{(S8)}$$

and

$$g(\mathrm{A}\,t^2) = 1/(1 + 4At^2)^{1/4}\ . \quad \text{(S9)}$$

$\varepsilon_c^{LSDA1}$ is the correlation energy of the uniform electron gas. Here we use the PW92 LSDA [12]. $\gamma = 0.031091$, $\beta(r_s) = 0.066725(1 + 0.1r_s)/(1 + 0.1778r_s)$ [8], and $\phi = \left[(1+\zeta)^{2/3} + (1-\zeta)^{2/3}\right]/2$. $\varepsilon_c^1$ differs from the original PBE correlation [6] only in the expressions for $\beta(r_s)$ and $g(At^2)$. The original PBE correlation has $\beta(r_s) = 0.066725$ and $g(At^2) = 1/(1 + At^2 + A^2t^4)$.

We design $\varepsilon_c^0$ in analogy to $\varepsilon_c^1$, realizing that for $\alpha$=0 only *s* and not *t* can arise:

$$\varepsilon_c^0 = (\varepsilon_c^{LDA0} + H_0)G_c(\zeta), \quad \text{(S10)}$$

where $G_c(\zeta) = \{1 - 2.3631[d_x(\zeta) - 1]\}(1 - \zeta^{12})$, (S11)

$$d_x(\zeta) = \left[(1+\zeta)^{4/3} + (1-\zeta)^{4/3}\right]/2, \quad \text{(S12)}$$

and $\varepsilon_c^{LDA0} = -b_{1c}/(1 + b_{2c} r_s^{1/2} + b_{3c} r_s)$. (S13)

$G_c(\zeta)$ was designed to make the correlation energy vanish for any one-electron density, and to make $F_{xc}(r_s \to \infty, \zeta, s = 0, \alpha = 0)$ independent of $\zeta$ for $0 \leq |\zeta| < 0.7$, a constraint [30] relevant to the atomization energy that was satisfied by TPSS and revTPSS. The exact exchange-correlation energy in the low-density limit is independent of $\zeta$. By achieving this for $s$=0 exactly at $\alpha = 1$ and as well as possible at α=0, our interpolation/extrapolation on $\alpha$ achieves it as well as possible for all α=0.

In analogy to $H_1$,

$$H_0 = b_{1c} \ln[1 + w_0(1 - g_\infty(\zeta = 0, s))], \quad \text{(S14)}$$

where $w_0 = exp[-\varepsilon_c^{LDA0}/b_{1c}] - 1$, (S15)

and

$$g_\infty(\zeta, s) = \lim_{r_s \to \infty} g(At^2) = 1/(1 + 4\chi_\infty s^2)^{1/4}. \quad \text{(S16)}$$

Here,

$\chi_\infty(\zeta) = \left(\frac{3\pi^2}{16}\right)^{\frac{2}{3}} \beta(r_s \to \infty)\phi/[c_x(\zeta) - f_0]$, $c_x(\zeta) = -(3/4\pi)(9\pi/4)^{1/3} d_x(\zeta)$, and $f_0 = -0.9$. At $\zeta = 0$, $\chi_\infty(\zeta = 0) = 0.128026$.

The parameters $b_{1c} = 0.0285764$, $b_{2c} = 0.0889$, and $b_{3c} = 0.125541$ are determined by the following procedure: In the high-density limit, $\varepsilon_c^0 = b_{1c} G_c(\zeta) \ln\{1 -$ $g_\infty(\zeta = 0, s)\exp(1)/[\exp(1) + 1]\}$ and $b_{1c} = 0.0285764$ by fitting to the correlation energy $E_c = -0.0467\ Ha$ of the high-density limit of the two-electron ion with the nucleus number $Z \to \infty$ [51]. $b_{3c} = 0.125541$ is determined by the lower bound on the exchange-correlation energies of 2-electron systems, $F_{xc} \leq 1.67082$ [47] . $b_{2c} = 0.0889$ is fitted to $E_{xc}$(He) = -1.068 Ha [15, 71]. The parameters in the interpolation/extrapolation function are $c_{1\mathrm{c}} = 0.64$, $c_{2\mathrm{c}} = 1.5$, and $d_\mathrm{c} = 0.7$.

The seven parameters ($c_{1x}$, $c_{2x}$, $d_x$, $k_1$, $c_{1c}$, $c_{2c}$, $d_c$) were determined in the following way: for a given $k_1$, we fitted 1) exactly to $\gamma_{x1} = -0.2259$ , $\gamma_{x2} = 0.2551$ , and $\gamma_{c1} = 0.0388$ , the large-Z asymptotic coefficients for exchange and correlation of rare-gas atoms of atomic number Z, 2) the binding energy curve of compressed $Ar_2$ (with a mean absolute error less than 1 kcal/mol for bond lengths R=1.6, 1.8. and 2.0 Å), 3) the jellium surface exchange-correlation energy at bulk density parameter $r_s$ = 4 Bohr within 5% of the QMC [63] value. Then we chose the parameter set with the maximum $k_1$, since the exact exchange energies for model metallic densities from Ref. [52] suggest that $k_1$ should not be too small.

SCAN is not fitted to any bonded system, but it predicts certain bonding properties very well: atomization energies, weak-interaction binding energies, and lattice constants of solids, but not the energy barriers to chemical reactions. We suggest that the first three properties fall naturally within the domain of a good semilocal functional, while the fourth requires a fully nonlocal approximation (e.g., a self-interaction correction to SCAN).

Note that there are recent empirical meta-GGAs [53-55].

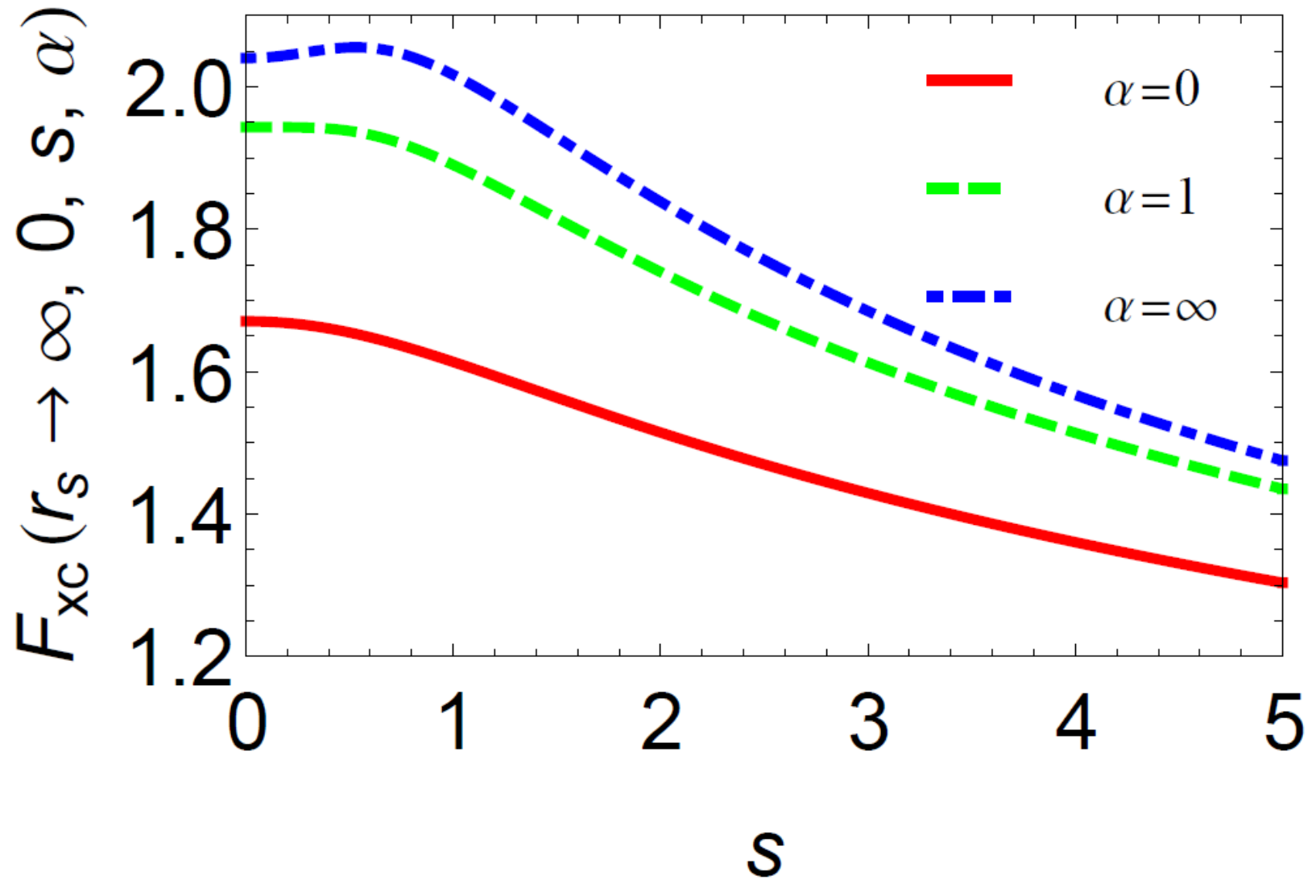

**(a)**

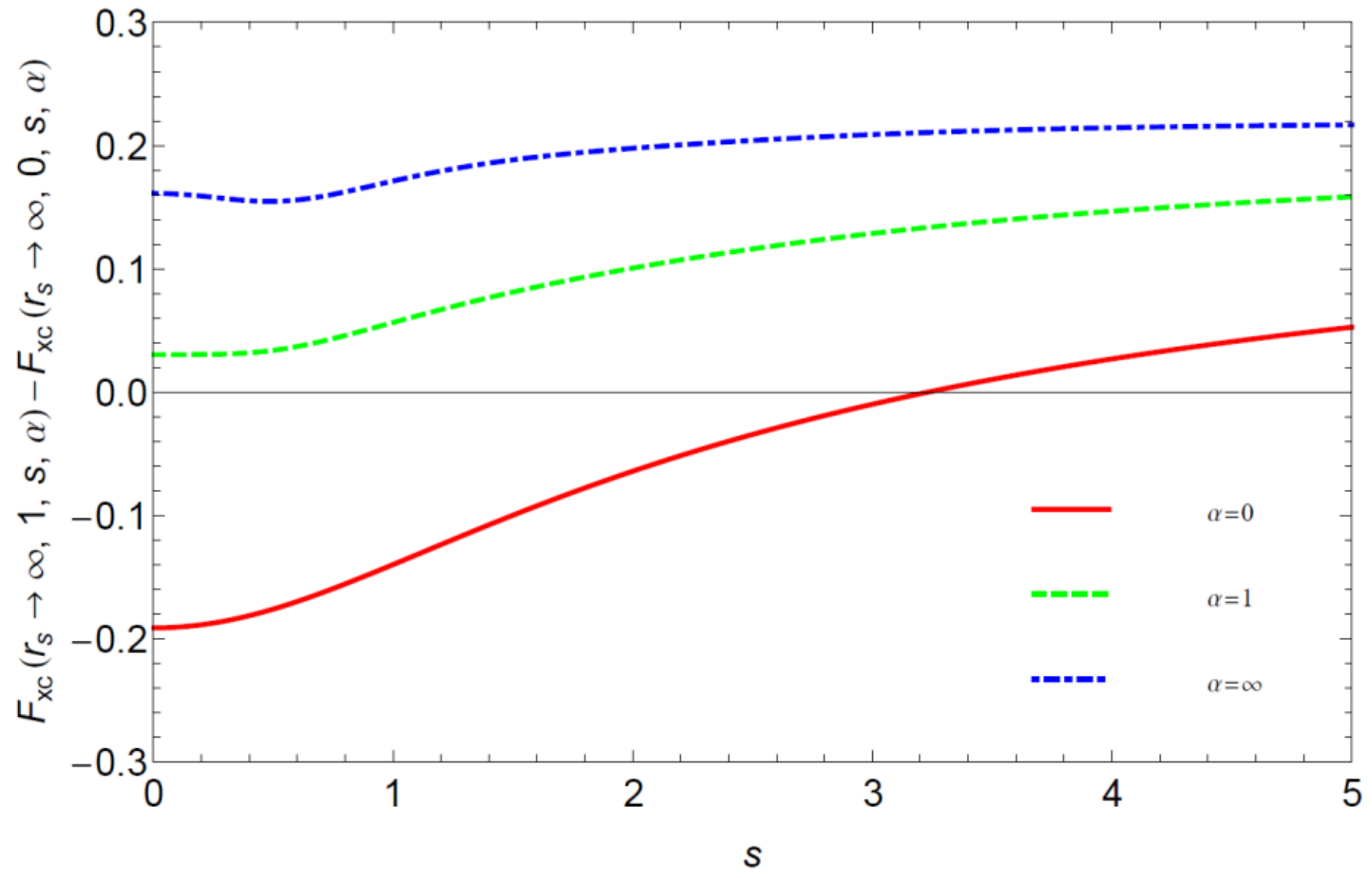

**(b)**

Figure S1. (a) The exchange-correlation enhancement factors in the low-density limit ($r_s \to \infty$) for spin-unpolarized densities ($\zeta = 0$). (b) The difference between the fully spin-polarized ($\zeta = 1$) and unpolarized ($\zeta = 0$) exchange-correlation enhancement factors in the low-density limit ($r_s \to \infty$).

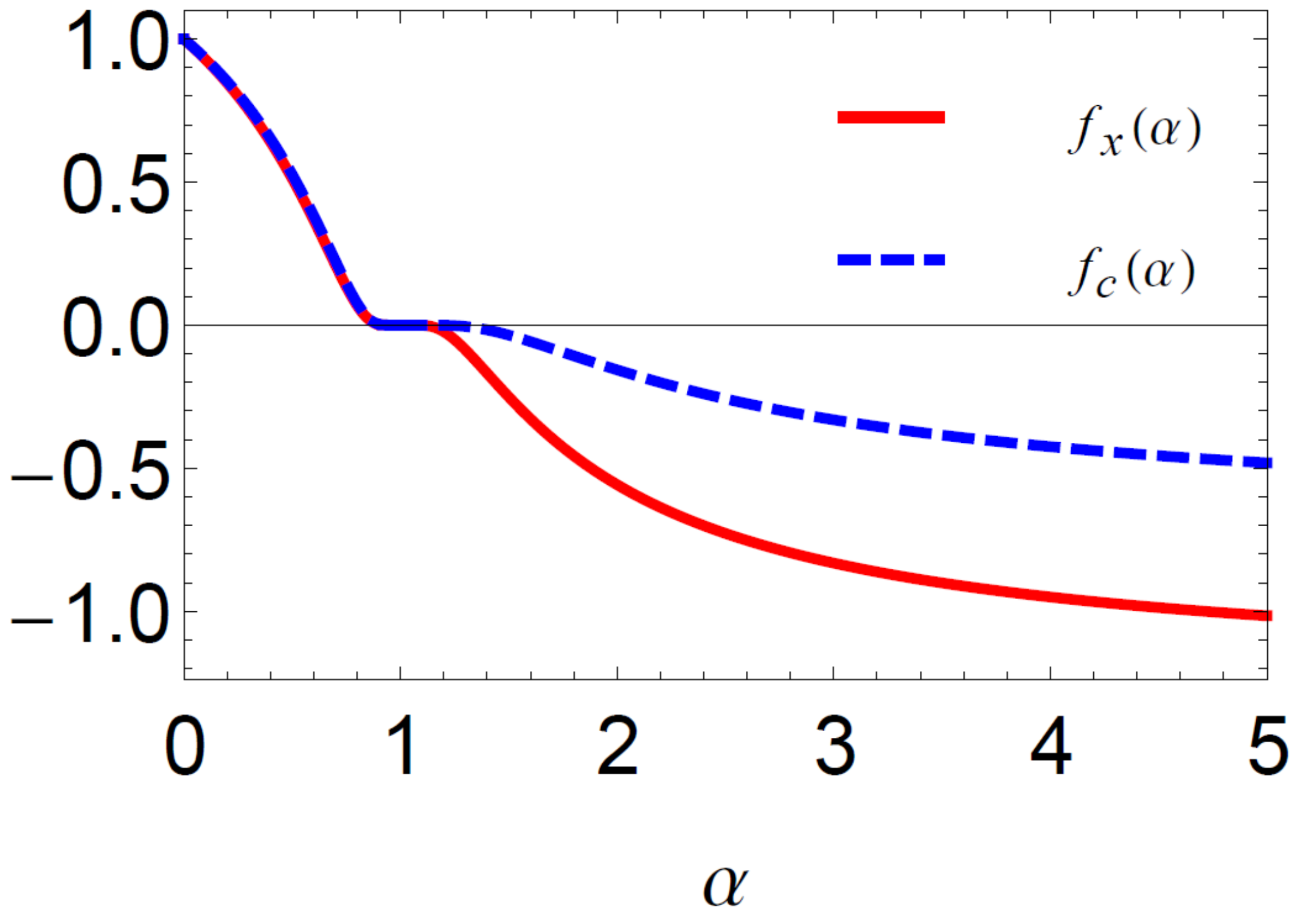

Figure S2. Interpolation/extrapolation functions $f_x(\alpha)$ and $f_c(\alpha)$.

Table SI. Values of parameters for exchange and the constraints and norms used to determine them. C and N denote constraints and norms, while the number or the letter in parenthesis indicates the specific one listed at the beginning of this Supplementary Material. Note $b_3 = 0.5$, not determined by the constraint C(4), was chosen to make the functional smooth in the variable space.

| $h_x^0$ | $a_1$ | $b_1$ | $b_2$ | $b_3$ | $b_4$ | $\mu_{AK}$ | $k_1$ | $c_{1x}$ | $c_{2x}$ | $d_x$ |
|---|---|---|---|---|---|---|---|---|---|---|
| 1.174 | 4.9479 | 0.1566 | 0.1208 | 0.5 | 0.1218 | 0.1234 | 0.065 | 0.667 | 0.8 | 1.24 |
| C(6) | N(c) | C(4) | | | | | N(b, d) | | | |

Table SII. Values of parameters for correlation and the constraints and norms used to determine them (parameters unchanged from the PBE GGA are not included here). C and N denote constraints and norms, while the number or the letter in parenthesis indicates the specific one listed at the beginning of this Supplementary Material.

| $b_{1c}$ | $b_{2c}$ | $b_{3c}$ | $c_{1c}$ | $c_{2c}$ | $d_c$ |
|---|---|---|---|---|---|
| 0.02858 | 0.0889 | 0.1255 | 0.64 | 1.5 | 0.7 |
| N(e) | N(d) | C(17) | N(b, d) | | |

Table SIII. Exact (or accurate) exchange [15], correlation [71, 72], and exchange-correlation energies of the rare-gas atoms. Unit: hartree.

| | Ne | Ar | Kr | Xe |
|---|---|---|---|---|
| $E_x$ | -12.108 | -30.188 | -93.890 | -179.200 |
| $E_c$ | -0.391 | -0.723 | -1.850 | -3.000 |
| Exc | -12.499 | -30.911 | -95.74 | -182.2 |

Table SIV. TPSS [7] and SCAN errors for the enthalpies of formation (kcal/mol) of the 223 molecules of the G3 test set [73], calculated selfconsistently from GAUSSIAN [68] using the 6-311+G(3df,3pd) basis set and standard geometries. For other functionals, see Ref. [25]. By construction, the error of the G3 enthalpy of formation is nearly equal in magnitude and opposite in sign to the error of the atomization energy. The experimental enthalpy of formation is also shown.

| Molecule | TPSS | SCAN | Expt. kcal/mol |
|---|---|---|---|
| ***G3-1 set*** | | | |
| LiH | -1.1 | 2.50 | 33.30 |
| BeH | -10.2 | -10.52 | 81.70 |
| CH | -3.4 | 2.11 | 142.80 |
| $CH_2(^3B_1)$ | -8.2 | -6.29 | 93.70 |
| $CH_2(^1A_1)$ | -0.5 | 5.44 | 102.30 |
| $CH_3$ | -6.3 | -4.72 | 35.00 |
| Methane($CH_4$) | -4.6 | 0.84 | -17.83 |
| NH | -6.7 | -1.16 | 85.20 |
| $NH_2$ | -6.1 | -2.63 | 45.10 |
| Ammonia($NH_3$) | -1.7 | 3.60 | -11.00 |
| OH | -0.6 | -2.54 | 9.40 |
| Water($H_2O$) | 3.5 | 2.77 | -57.80 |
| Hydrogenfluoride(HF) | 1.3 | 3.16 | -65.10 |
| $SiH_2(^1A_1)$ | -5.3 | 2.29 | 65.20 |
| $SiH_2(^3B_1)$ | -10.7 | -7.59 | 86.20 |
| $SiH_3$ | -11.8 | -4.97 | 47.90 |
| Silane ($SiH_4$) | -11.6 | -1.15 | 8.20 |
| $PH_2$ | -8.8 | -4.41 | 33.10 |
| $PH_3$ | -8.0 | -0.69 | 1.30 |
| Hydrogen sulfide ($H_2S$) | -3.0 | -0.14 | -4.90 |
| Hidrogen chloride (HCl) | -1.1 | 0.16 | -22.10 |
| $Li_2$ | 1.2 | 5.95 | 51.60 |
| LiF | 1.2 | 3.39 | -80.10 |
| Acetylene($C_2H_2$) | 0.7 | 4.20 | 54.35 |
| Ethylene($H_2C{=}CH_2$) | -4.3 | 0.92 | 12.52 |
| Ethane($H_3C-CH_3$) | -6.1 | -0.56 | -20.10 |
| CN | -1.6 | 3.93 | 104.90 |
| Hydrogencyanide(HCN) | -1.0 | 5.38 | 31.50 |
| CO | 4.2 | 4.08 | -26.40 |
| HCO | -4.9 | -4.28 | 10.00 |
| Formaldehyde($H_2C{=}O$) | -3.1 | -0.48 | -26.00 |
| Methanol ($CH_3$-OH) | -2.6 | -0.87 | -48.00 |
| $N_2$ | 0.9 | 9.79 | 0.00 |
| Hydrazine($H_2N-NH_2$) | -5.0 | 3.97 | 22.75 |
| NO | -4.1 | 1.22 | 21.60 |
| $O_2$ | -6.8 | -7.25 | 0.00 |
| Hydrogenperoxide(HO-OH) | -0.2 | 0.74 | -32.50 |
| $F_2$ | -6.7 | 2.13 | 0.00 |
| Carbon dioxide ($CO_2$) | -1.8 | -5.51 | -94.05 |
| $Na_2$ | -2.4 | 2.79 | 34.00 |
| $Si_2$ | -3.8 | -2.63 | 141.00 |
| $P_2$ | 0.0 | 3.97 | 34.30 |
| $S_2$ | -7.1 | -7.01 | 30.70 |
| $Cl_2$ | -3.2 | 0.32 | 0.00 |
| NaCl | 0.8 | 0.05 | -43.60 |
| Silicon monoxide (SiO) | 5.6 | 5.56 | -24.60 |
| CS | 1.5 | 3.76 | 66.90 |
| SO | -4.9 | -6.38 | 1.20 |
| ClO | -7.5 | -4.93 | 24.20 |
| Chlorine monofluoride (FCl) | -5.5 | 0.29 | -13.20 |
| $Si_2H_6$ | -17.9 | -4.79 | 19.10 |
| Methyl chloride ($CH_3Cl$) | -5.5 | -1.99 | -19.60 |
| Methanethiol ($H_3CSH$) | -5.3 | -1.31 | -5.50 |
| Hypochlorous acid (HOCl) | -2.5 | -0.15 | -17.80 |
| Sulfur dioxide ($SO_2$) | 1.5 | -1.07 | -71.00 |
| ***G3-2 set*** | | | |
| $BF_3$ | **3.9** | -2.19 | -271.40 |
| $BCl_3$ | -2.1 | -13.07 | -96.30 |
| $AlF_3$ | 9.3 | 5.72 | -289.00 |
| $AlCl_3$ | -0.1 | -9.78 | -139.70 |
| Carbon tetrafuoride ($CF_4$) | -4.2 | -8.36 | -223.00 |
| Carbon tetrachloride ($CCl_4$) | -2.5 | -8.06 | -22.90 |
| Carbon oxide sulfide (COS) | -6.0 | -7.73 | -33.10 |
| Carbon bisulfide ($CS_2$) | -8.5 | -8.80 | 28.00 |
| Carbonic difluoride ($COF_2$) | -0.3 | -3.53 | -149.10 |
| Silicon tertrafluoride ($SiF_4$) | 16.2 | 10.73 | -386.00 |
| Silicon tetrachloride ($SiCl_4$) | 3.4 | -8.23 | -158.40 |
| Dinitrogen monoxide ($N_2O$) | -12.0 | -1.92 | 19.60 |
| Nitrogen chloride oxide (ClNO) | -13.4 | -6.16 | 12.40 |
| Nitrogen trifluoride ($NF_3$) | -19.4 | -8.40 | -31.60 |
| $PF_3$ | 1.8 | 3.85 | -229.10 |
| $O_3$ | -9.0 | -2.22 | 34.10 |
| $F_2O$ | -14.3 | -4.53 | 5.90 |
| Chlorine trifluoride ($ClF_3$) | -22.9 | -13.31 | -38.00 |
| Ethene, tetrafluoro- ($F_2C=CF_2$) | -14.8 | -16.58 | -157.40 |
| Ethene, tetrachloro- ($C_2Cl_4$) | -6.5 | -13.09 | -3.00 |
| Acetonitrile, trifluoro- ($CF_3CN$) | -5.5 | -5.39 | -118.40 |
| Propyne($C_3H_4$) | -2.6 | 0.95 | 44.20 |
| Allene($C_3H_4$) | -7.0 | -2.93 | 45.50 |
| Cyclopropene($C_3H_4$) | -5.2 | 0.58 | 66.20 |
| Propylene($C_3H_6$) | -5.6 | -1.12 | 4.80 |
| Cyclopropane($C_3H_6$) | -7.4 | -2.94 | 12.70 |
| Propane($C_3H_8$) | -6.6 | -1.88 | -25.00 |
| Trans-1,3-butadiene ($C_4H_6$) | -5.8 | -2.22 | 26.30 |
| Dimethylacetylene ($C_4H_6$) | -4.8 | -0.98 | 34.80 |
| Methylenecyclopropane ($C_4H_6$) | -10.7 | -6.75 | 47.90 |
| Bicyclo[1.1.0]butane ($C_4H_6$) | -7.7 | -4.03 | 51.90 |
| Cyclobutene ($C_4H_6$) | -5.1 | -2.19 | 37.40 |
| Cyclobutane ($C_4H_8$) | -7.1 | -4.30 | 6.80 |
| Isobutene($C_4H_8$) | -5.6 | -2.58 | -4.00 |
| Trans-butane($C_4H_{10}$) | -6.8 | -2.95 | -30.00 |
| Isobutane($C_4H_{10}$) | -5.6 | -2.58 | -32.10 |
| Spiropentane($C_5H_8$) | -10.8 | -7.69 | 44.30 |
| Benzene($C_6H_6$) | -5.5 | -8.88 | 19.70 |
| Difluoromethane($CH_2F_2$) | -7.3 | -3.82 | -108.10 |
| Trifluoromethane($CHF_3$) | -6.1 | -5.96 | -166.60 |
| $CH_2Cl_2$ | -5.2 | -4.21 | -22.80 |
| $CHCl_3$ | -4.0 | -6.16 | -24.70 |
| Methylamine($H_3C$-$NH_2$) | -5.1 | 1.72 | -5.50 |
| Acetonitrile($CH_3$-CN) | -4.1 | 2.24 | 18.00 |
| Nitromethane($CH_3$-$NO_2$) | -13.3 | -8.36 | -17.80 |
| Methyl nitrite($CH_3$-O-N=O) | -13.3 | -6.81 | -15.90 |
| Methyl silane ($CH_3SiH_3$) | -10.0 | -0.94 | -7.00 |
| Formic acid (HCOOH) | -0.8 | -4.54 | -90.50 |
| Methyl formate ($HCOOCH_3$) | -6.4 | -8.53 | -85.00 |
| Acetamide ($CH_3CONH_2$) | -4.3 | -4.78 | -57.00 |
| Aziridine ($C_2H_4NH$) | -9.1 | -1.53 | 30.20 |
| Cyanogen (NCCN) | -3.5 | 5.25 | 73.30 |
| Dimethylamine ($(CH_3)_2NH$ ) | -7.6 | -0.35 | -4.40 |
| Trans ethylamine ($CH_3CH_2NH_2$) | -6.6 | -0.76 | -11.30 |
| Ketene ($CH_2CO$) | -6.0 | -6.20 | -11.40 |
| Oxirane ($C_2H_4O$) | -8.2 | -4.89 | -12.60 |
| Acetaldehyde ($CH_3CHO$) | -4.4 | -3.05 | -39.70 |
| Glyoxal (HCOCOH) | -2.8 | -4.35 | -50.70 |
| Ethanol ($CH_3CH_2OH$) | -2.9 | -2.21 | -56.20 |
| Dimethylether ($CH_3OCH_3$) | -7.1 | -3.98 | -44.00 |
| Thiirane ($C_2H_4S$) | -8.4 | -4.59 | 19.60 |
| Dimethyl sulfoxide ($(CH_3)_2SO$) | -5.6 | -5.54 | -36.20 |
| Ethanethiol ($C_2H_5SH$) | -5.5 | -2.33 | -11.10 |
| Dimethyl sulfide ($CH_3SCH_3$) | -7.1 | -2.79 | -8.90 |
| Vinyl fluoride ($CH_2$=CHF) | -7.3 | -4.16 | -33.2 |
| Ethyl chloride ($C_2H_5Cl$) | -6.2 | -3.68 | -26.80 |
| Vinyl chloride (CH2=CHCl) | -8.8 | -6.54 | 8.90 |
| Acrylonitrile($CH_2$=CHCN) | -2.7 | 3.26 | 43.2 |
| Acetone ($CH_3COCH_3$) | -4.7 | -5.02 | -51.9 |
| Acetic acid ($CH_3COOH$) | -0.9 | -6.31 | -103.4 |
| Acetyl fluoride ($CH_3COF$) | -4.8 | -6.51 | -105.7 |
| $CH_3COCl$ (acetyl chloride) | -5.9 | -8.27 | -58.00 |
| $CH_3CH_2CH_2Cl$ (propyl chloride) | -6.8 | -5.06 | -31.50 |
| Isopropanol $(CH_3)_2CHOH$) | -2.5 | -3.61 | -65.2 |
| Methyl ethyl ether ($C_2H_5OCH_3$) | -7.8 | -5.74 | -51.7 |
| Trimethylamine ($(CH_3)_3N$) | -9.6 | -2.76 | -5.7 |
| Furan ($C_4H_4O$) | -5.8 | -9.03 | -8.3 |
| $C_4H_4S$ (thiophene) | -5.1 | -7.44 | 27.50 |
| Pyrrole ($C_4H_5N$) | -7.3 | -6.49 | 25.9 |
| Pyridine ($C_5H_5N$) | -8.7 | -8.74 | 33.6 |
| $H_2$ | -3.2 | -3.26 | 0.00 |
| HS | -2.8 | -2.06 | 34.20 |
| CCH | 0.1 | 0.18 | 135.10 |
| $C_2H_3$ ($^2A'$) | -7.5 | -5.74 | 71.60 |
| $CH_3CO$ ($^2A'$) | -7.1 | -7.10 | -2.40 |
| $H_2COH$ ($^2A$) | -5.0 | -5.95 | -4.10 |
| $CH_3O$ Cs($^2A'$) | -8.1 | -7.50 | 4.10 |
| $CH_3CH_2O$ ($^2A''$) | -9.9 | -10.09 | -3.70 |
| $CH_3S$ ($^2A'$) | -6.9 | -4.81 | 29.80 |
| $C_2H_5$ ($^2A'$) | -8.6 | -6.78 | 28.90 |
| $(CH_3)_2CH$ ($^2A'$) | -10.3 | -8.92 | 21.50 |
| $(CH_3)_3C$ (t-butyl radical) | -10.1 | -9.67 | 12.30 |
| $NO_2$ | -14.3 | -9.34 | 7.90 |
| ***G3-3 set*** | | | |
| Methyl allene ($C_4H_6$) | -8.0 | -4.11 | 38.8 |
| Isoprene ($C_5H_8$) | -5.3 | -3.35 | 18 |
| Cyclopentane ($C_5H_{10}$) | -4.9 | -5.00 | -18.3 |
| n-Pentane ($C_5H_{12}$) | -6.9 | -3.87 | -35.1 |
| Neo pentane ($C_5H_{12}$) | -3.7 | -3.04 | -40.2 |
| 1,3 Cyclohexadiene ($C_6H_8$) | -3.1 | -5.64 | 25.4 |
| 1,4 Cyclohexadiene ($C_6H_8$) | -2.2 | -4.96 | 25 |
| Cyclohexane ($C_6H_{12}$) | -2.7 | -5.87 | -29.5 |
| n-Hexane ($C_6H_{14}$) | -7.3 | -5.14 | -39.9 |
| 3-Methyl pentane ($C_6H_{14}$) | -5.2 | -4.04 | -41.1 |
| Toluene ($C_6H_5CH_3$ ) | -5.6 | -10.20 | 12 |
| n-Heptane ($C_7H_{16}$) | -7.5 | -6.15 | -44.9 |
| Cyclooctatetraene ($C_8H_8$) | -2.0 | -6.98 | 70.7 |
| n-Octane ($C_8H_{18}$) | -7.7 | -7.24 | -49.9 |
| Naphthalene ($C_{10}H_8$) | -5.6 | -18.50 | 35.9 |
| Azulene ($C_{10}H_8$) | -8.3 | -19.12 | 69.10 |
| Acetic acid methyl ester ($CH_3COOCH_3$) | -5.4 | -9.18 | -98.40 |
| t-Butanol $(CH_3)_3COH$ | -0.9 | -4.41 | -74.70 |
| Aniline ($C_6H_5NH_2$) | -6.5 | -10.91 | 20.80 |
| Phenol ($C_6H_5OH$) | -2.6 | -11.98 | -23.00 |
| Divinyl ether ($C_4H_6O$) | -7.0 | -6.85 | -3.30 |
| Tetrahydrofuran ($C_4H_8O$) | -5.0 | -6.92 | -44.00 |
| Cyclopentanone($C_5H_8O$) | -4.0 | -9.31 | -45.90 |
| Benzoquinone($C_6H_4O_2$) | -1.3 | -12.44 | -29.40 |
| Pyrimidine($C_4H_4N_2$) | -12.7 | -9.73 | 46.80 |
| Dimethyl sulphone ($C_2H_6O_2S$) | 0.2 | -7.89 | -89.20 |
| Chlorobenzene ($C_6H_5Cl$) | -5.8 | -12.65 | 12.40 |
| Butanedinitrile(NºC-$CH_2$-$CH_2$-CºN) | -2.6 | 3.94 | 50.10 |
| Pyrazine($C_4H_4N_2$) | -9.1 | -5.67 | 46.90 |
| Acetyl acetylene ($CH_3$-C(=O)-CºCH) | 0.2 | -1.25 | 15.60 |
| Crotonaldehyde ($CH_3$-CH=CH-CHO) | -7.2 | -7.41 | -24.00 |
| Acetic anhydride ($CH_3$-C(=O)-O-C(=O)-$CH_3$) | -5.0 | -14.97 | -136.80 |
| 2,5-Dihydrothiophene ($C_4H_6S$) | -5.1 | -6.06 | 20.80 |
| Isobutane nitrile($(CH_3)_2CH$-CN) | -2.7 | 1.88 | 5.60 |
| Methyl ethyl ketone($CH_3$-CO-$CH_2$-$CH_3$) | -4.9 | -6.22 | -57.10 |
| Isobutanal($(CH_3)_2CH$-CHO) | -3.4 | -4.04 | -51.60 |
| 1,4-Dioxane($C_4H_8O_2$) | -5.0 | -11.41 | -75.50 |
| Tetrahydrothiophene ($C_4H_8S$) | -4.3 | -5.36 | -8.20 |
| t-Butyl chloride ($(CH_3)_3C$-Cl) | -4.7 | -6.38 | -43.50 |
| n-Butyl chloride ($CH_3$-$CH_2$-$CH_2$-$CH_2$-Cl) | -6.5 | -5.57 | -37.00 |
| Tetrahydropyrrole($C_4H_8NH$) | -6.3 | -3.97 | -0.80 |
| Nitro-s-butane ($CH_3$-$CH_2$-CH($CH_3$)-$NO_2$) | -12.2 | -11.07 | -39.10 |
| Diethyl ether($CH_3$-$CH_2$-O-$CH_2$-$CH_3$) | -7.6 | -6.57 | -60.30 |
| Dimethyl acetal($CH_3$-CH($OCH_3$)$_2$) | -6.8 | -9.60 | -93.10 |
| t-Butanethiol ($(CH_3)_3C$-SH) | -3.6 | -4.43 | -26.20 |
| Diethyl disulfide ($CH_3$-$CH_2$-S-S-$CH_2$-$CH_3$) | -8.1 | -6.38 | -17.90 |
| t-Butylamine ($(CH_3)_3C$-$NH_2$) | -3.3 | -1.72 | -28.90 |
| Tetramethylsilane ($Si(CH_3)_4$) | -3.0 | 0.15 | -55.70 |
| 2-Methyl thiophene ($C_5H_6S$) | -5.7 | -8.98 | 20.00 |
| N-methyl pyrrole (cyc-$C_4H_4N$-$CH_3$) | -9.1 | -7.98 | 24.60 |
| Tetrahydropyran($C_5H_{10}O$) | -3.7 | -8.50 | -53.40 |
| Diethyl ketone ($CH_3$-$CH_2$-CO-$CH_2$-$CH_3$) | -5.8 | -8.08 | -61.60 |
| Isopropyl acetate ($CH_3$-C(=O)-O-CH($CH_3$)$_2$) | -4.9 | -11.53 | -115.10 |
| Tetrahydrothiopyran ($C_5H_{10}S$) | -3.3 | -6.90 | -15.20 |
| Piperidine(cyc-$C_5H_{10}NH$) | -4.1 | -4.81 | -11.30 |
| t-Butyl methyl ether($(CH_3)_3C$-O-$CH_3$) | -4.6 | -6.86 | -67.80 |
| 1,3-Difluorobenzene($C_6H_4F_2$) | -9.4 | -17.85 | -73.90 |
| 1,4-Difluorobenzene($C_6H_4F_2$) | -9.3 | -17.53 | -73.30 |
| Fluorobenzene ($C_6H_5F$) | -7.2 | -13.08 | -27.70 |
| Di-isopropyl ether ($(CH_3)_2CH$-O-CH($CH_3$)$_2$) | -5.6 | -8.23 | -76.30 |
| $PF_5$ | 7.5 | 3.75 | -381.10 |
| $SF_6$ | -3.8 | -10.36 | -291.70 |
| $P_4$ | -9.9 | -4.97 | 14.10 |
| $SO_3$ | 1.9 | -5.57 | -94.60 |
| $SCl_2$ | -5.6 | -3.81 | -4.20 |
| $POCl_3$ | 0.8 | -9.00 | -133.80 |
| $PCl_5$ | -7.3 | -16.91 | -86.10 |
| $Cl_2O_2S$ | -0.6 | -10.42 | -84.80 |
| $PCl_3$ | -4.7 | -7.31 | -69.00 |
| $Cl_2S_2$ | -12.8 | -12.09 | -4.00 |
| $SiCl_2$ singlet | -0.9 | -3.04 | -40.30 |
| $CF_3Cl$ | -5.9 | -10.36 | -169.50 |
| Ethane,-hexafluoro- ($C_2F_6$) | -6.6 | -16.53 | -321.30 |
| $CF_3$ | -10.8 | -12.32 | -111.30 |
| $C_6H_5$ (phenyl radical) | -9.6 | -15.93 | 81.20 |

Table SV. Error statistics of SCAN and other functionals for the G3-1, G3-2, and G3-3, subsets of G3 [73] that contain molecules increasing in size on average from -1 to -3. ME and MAE are mean error and mean absolute error.

| | | LSDA | BLYP | PBEsol | PBE | TPSS | M06L | SCAN |
|---|---|---|---|---|---|---|---|---|
| G3-1 | ME | -36.3 | -2.9 | -16.9 | -6.7 | -3.7 | -0.4 | -0.2 |
| | MAE | 36.3 | 4.8 | 17.2 | 8.2 | 4.5 | 3.5 | 3.3 |
| G3-2 | ME | -111.1 | 0.8 | -54.8 | -21.6 | -6.1 | -2.4 | -4.6 |
| | MAE | 111.1 | 8.7 | 54.9 | 22.0 | 6.9 | 5.2 | 5.4 |
| G3-3 | ME | -196.6 | 12.4 | -94.2 | -32.8 | -5.2 | -1.4 | -7.8 |
| | MAE | 196.6 | 13.9 | 94.2 | 32.8 | 5.5 | 6.4 | 8.0 |
| G3 | ME | -121.4 | 3.8 | -58.7 | -21.7 | -5.2 | -1.6 | -4.6 |
| | MAE | 121.4 | 9.5 | 58.5 | 22.2 | 5.8 | 5.2 | 5.7 |

Table SVI. Errors for 46 hydrocarbon molecules of the $G3^{HC}$ set [73], after an empirical correction to the energy of the carbon atom is taken into account. This additive correction is –n($\delta E$), where n is the number of C atoms in a molecule. $\delta E$ is the average error per C atom over the hydrocarbon molecules. Note that the SCAN errors after this correction are the smallest of all the tested functionals. ME* and MAE* are mean error and mean absolute error before the correction.

| | LSDA | BLYP | PBEsol | PBE | TPSS | M06L | SCAN |
|---|---|---|---|---|---|---|---|
| CH | 26.6 | -4.4 | 12.9 | 3.8 | -1.6 | 1.7 | 3.0 |
| $CH_2(^3B_1)$ | 12.9 | -2.4 | 5.0 | 0.5 | -6.4 | -1.8 | -5.4 |
| $CH_2(^1A_1)$ | 17.1 | -1.9 | 11.7 | 6.4 | 1.3 | 2.0 | 6.4 |
| $CH_3$ | 3.7 | -2.2 | 3.7 | 2.2 | -4.5 | 0.7 | -3.8 |
| Methane($CH_4$) | -6.6 | 0.6 | 1.9 | 4.9 | -2.8 | 2.2 | 1.8 |
| Acetylene($C_2H_2$) | 16.1 | -4.7 | 7.0 | 0.2 | 4.4 | 0.2 | 6.1 |
| Ethylene($H_2C{=}CH_2$) | 1.5 | -2.7 | 2.3 | 1.2 | -0.6 | 1.8 | 2.8 |
| Ethane($H_3C$-$CH_3$) | -11.1 | 2.5 | 0.0 | 4.8 | -2.4 | 2.8 | 1.3 |
| Propyne($C_3H_4$) | 8.7 | -4.6 | 2.4 | -2.0 | 2.9 | -2.1 | 3.7 |
| Allene($C_3H_4$) | 3.6 | -9.1 | -2.5 | -6.7 | -1.5 | -4.1 | -0.1 |
| Cyclopropene($C_3H_4$) | 3.6 | -2.1 | -2.5 | -4.9 | 0.3 | -5.6 | 3.4 |
| Propylene($C_3H_6$) | -4.4 | -0.7 | -0.5 | 0.8 | -0.1 | 1.7 | 1.7 |
| Cyclopropane($C_3H_6$) | -9.6 | 2.7 | -5.1 | -1.4 | -1.9 | -2.8 | -0.2 |
| Propane($C_3H_8$) | -15.6 | 5.3 | -1.6 | 5.4 | -1.1 | 4.0 | 0.9 |
| Trans-1,3-butadiene ($C_4H_6$) | 2.3 | -4.5 | -1.1 | -3.6 | 1.6 | 0.1 | 1.5 |
| Dimethylacetylene ($C_4H_6$) | 2.5 | -3.1 | -0.9 | -3.1 | 2.6 | -2.8 | 2.7 |
| Methylenecyclopropane ($C_4H_6$) | -6.9 | -4.5 | -9.6 | -9.5 | -3.3 | -8.2 | -3.0 |
| Bicyclo[1.1.0]butane ($C_4H_6$) | -7.2 | 4.5 | -9.2 | -6.6 | -0.3 | -8.6 | -0.3 |
| Cyclobutene ($C_4H_6$) | -2.3 | 2.4 | -5.0 | -4.0 | 2.3 | -0.9 | 1.5 |
| Cyclobutane ($C_4H_8$) | -14.5 | 6.7 | -7.0 | -0.4 | 0.3 | 1.1 | -0.6 |
| Isobutene($C_4H_8$) | -9.6 | 2.7 | -2.3 | 1.5 | 1.8 | 2.7 | 1.1 |
| Trans-butane($C_4H_{10}$) | -20.1 | 8.4 | -3.0 | 6.1 | 0.6 | 5.4 | 0.8 |
| Isobutane($C_4H_{10}$) | -20.0 | 9.7 | -2.5 | 7.1 | 1.8 | 6.3 | 1.1 |
| Spiropentane($C_5H_8$) | -14.5 | 4.0 | -13.5 | -8.8 | -1.6 | -8.9 | -3.0 |
| Benzene($C_6H_6$) | 5.0 | -5.1 | -9.5 | -14.0 | 5.6 | -5.9 | -3.3 |
| CCH | 26.8 | -6.1 | 8.6 | -1.9 | 3.8 | -5.5 | 2.0 |
| $C_2H_3$ ($^2A'$) | 7.9 | -7.4 | 0.9 | -3.8 | -3.8 | -3.5 | -3.9 |
| $C_2H_5$ ($^2A'$) | -3.3 | -1.6 | -0.1 | 0.7 | -4.9 | -0.5 | -4.9 |
| $(CH_3)_2CH$ ($^2A'$) | -9.9 | -0.4 | -3.6 | -0.3 | -4.8 | -1.2 | -6.1 |
| $(CH_3)_3C$ (t-butyl radical) | -15.3 | 3.1 | -5.4 | 0.5 | -2.7 | -0.1 | -6.0 |
| Methyl allene ($C_4H_6$) | -1.0 | -6.7 | -4.3 | -6.5 | -0.6 | -3.2 | -0.4 |
| Isoprene ($C_5H_8$) | -3.1 | -0.3 | -2.7 | -2.4 | 3.9 | 1.0 | 1.3 |
| Cyclopentane ($C_5H_{10}$) | -19.3 | 12.1 | -7.7 | 2.0 | 4.3 | 5.5 | -0.4 |
| n-Pentane ($C_5H_{12}$) | -24.3 | 11.6 | -4.3 | 7.1 | 2.3 | 6.9 | 0.8 |
| Neo pentane ($C_5H_{12}$) | -24.1 | 15.3 | -2.8 | 9.7 | 5.5 | 8.7 | 1.6 |
| 1,3 Cyclohexadiene ($C_6H_8$) | -3.1 | 3.1 | -7.9 | -6.7 | 8.0 | 1.0 | -0.1 |
| 1,4 Cyclohexadiene ($C_6H_8$) | -2.5 | 4.0 | -7.1 | -6.0 | 8.9 | 1.3 | 0.6 |
| Cyclohexane ($C_6H_{12}$) | -24.1 | 17.5 | -8.2 | 4.7 | 8.4 | 7.9 | -0.3 |
| n-Hexane ($C_6H_{14}$) | -29.0 | 14.6 | -5.9 | 7.8 | 3.8 | 8.2 | 0.4 |
| 3-Methyl pentane ($C_6H_{14}$) | -28.9 | 17.2 | -4.9 | 9.6 | 5.9 | 9.1 | 1.5 |
| Toluene ($C_6H_5CH_3$ ) | -35.6 | 0.7 | -27.3 | -18.2 | 5.5 | -5.6 | -4.6 |
| n-Heptane ($C_7H_{16}$) | -33.3 | 17.7 | -7.3 | 8.6 | 5.4 | 9.6 | 0.4 |
| Cyclooctatetraene ($C_8H_8$) | 9.7 | -3.1 | -8.8 | -14.7 | 12.8 | -2.9 | 0.5 |
| n-Octane ($C_8H_{18}$) | -37.7 | 20.9 | -8.7 | 9.4 | 7.1 | 11.0 | 0.2 |
| Naphthalene ($C_{10}H_8$) | 8.7 | -6.6 | -20.9 | -28.6 | 12.9 | -12.9 | -9.2 |
| Azulene ($C_{10}H_8$) | 7.1 | -9.8 | -23.3 | -31.6 | 10.2 | -14.2 | -9.8 |
| $C_6H_5$ (phenyl radical) | 11.2 | -10.4 | -11.5 | -19.7 | 1.5 | -12.4 | -10.4 |
| ME | -5.6 | 1.8 | -4.1 | -2.1 | 1.9 | -0.2 | -0.6 |
| MAE | 13.0 | 6.2 | 6.5 | 6.6 | 3.8 | 4.6 | 2.7 |
| $\delta E$ | -35.5 | 2.5 | -15.9 | -4.8 | -1.8 | -0.6 | -0.9 |
| ME* | -154.9 | 12.4 | -71.0 | -22.4 | -5.8 | -2.8 | -4.5 |
| MAE* | 154.9 | 12.7 | 71.0 | 22.5 | 5.9 | 5.0 | 5.1 |

Table SVII. SCAN errors for the BH76 barrier heights to chemical reactions [76]. For other functionals, see Ref [34]. $V_f^{\neq}$ is the forward reaction barrier, and$V_r^{\neq}$ is the backward reaction barrier. Accurate barriers are taken from Ref. [77]. Unit: kcal/mol.

| Reaction | $V_f^{\neq}$ | | $V_r^{\neq}$ | |
|---|---|---|---|---|
| A+BC→AB+C | accurate | SCAN | accurate | SCAN |
| H + HCl → $H_2$ + Cl | 5.7 | -7.06 | 8.7 | -8.81 |
| OH + $H_2$ → H + $H_2O$ | 5.7 | -7.91 | 21.2 | -10.12 |
| $CH_3$ + $H_2$ → H + $CH_4$ | 12.1 | -4.88 | 15.3 | -8.49 |
| OH + $CH_4$ → $CH_3$ + $H_2O$ | 6.7 | -8.68 | 19.6 | -7.88 |
| H + $H_2$ → $H_2$ + H | 9.6 | -7.26 | 9.6 | -7.26 |
| OH + $NH_3$ → $H_2O$ + $NH_2$ | 3.2 | -10.83 | 12.7 | -9.50 |
| HCl + $CH_3$ → Cl + $CH_4$ | 1.7 | -4.97 | 7.9 | -10.34 |
| $OH + C_2H_6 \rightarrow H_2O + C_2H_5$ | 3.4 | -8.57 | 19.9 | -6.94 |
| $F + H_2 \rightarrow HF + H$ | 1.8 | -9.70 | 33.4 | -11.01 |
| $O + CH_4 \rightarrow OH + CH_3$ | 13.7 | -12.28 | 8.1 | -4.81 |
| $H + PH_3 \rightarrow PH_2 + H_2$ | 3.1 | -6.39 | 23.2 | -4.19 |
| $H + HO \rightarrow H_2 + O$ | 10.7 | -7.43 | 13.1 | -11.29 |
| $H + H_2S \rightarrow H_2 + HS$ | 3.5 | -6.30 | 17.3 | -6.22 |
| $O + HCl \rightarrow OH + Cl$ | 9.8 | -13.81 | 10.4 | -11.71 |
| $NH_2 + CH_3 \rightarrow CH_4 + NH$ | 8.0 | -3.52 | 22.4 | -10.52 |
| $NH_2 + C_2H_5 \rightarrow C_2H_6 + NH$ | 7.5 | -1.55 | 18.3 | -9.36 |
| $C_2H_6 + NH_2 \rightarrow NH_3 + C_2H_5$ | 10.4 | -5.81 | 17.4 | -5.51 |
| $NH_2 + CH_4 \rightarrow CH_3 + NH_3$ | 14.5 | -7.04 | 17.8 | -7.47 |
| *s-trans cis*-$C_5H_8$ → *s-trans cis*-$C_5H_8$ | 38.4 | -4.77 | 38.4 | -4.77 |
| $H + N_2O \rightarrow OH + N_2$ | 18.14 | -8.81 | 83.22 | -18.65 |
| $H + FH \rightarrow HF + H$ | 42.18 | -13.45 | 42.18 | -13.45 |
| $H + ClH \rightarrow HCl + H$ | 18.00 | -8.72 | 18.00 | -8.72 |
| $H + FCH_3 \rightarrow HF + CH_3$ | 30.38 | -10.44 | 57.02 | -10.71 |
| $H + F_2 \rightarrow HF + F$ | 2.27 | -13.86 | 106.18 | -16.29 |
| $CH_3 + FCl \rightarrow CH_3F + Cl$ | 7.43 | -12.26 | 61.01 | -15.43 |
| $F^- + CH_3F \rightarrow FCH_3 + F^-$ | -0.34 | -8.00 | -0.34 | -8.00 |
| $F^- \cdots CH_3F \rightarrow FCH_3 \cdots F^-$ | 13.38 | -5.34 | 13.38 | -5.34 |
| $Cl^- + CH_3Cl \rightarrow ClCH_3 + Cl^-$ | 3.10 | -8.20 | 3.10 | -8.20 |
| $Cl^- \cdots CH_3Cl \rightarrow ClCH_3 \cdots Cl^-$ | 13.61 | -6.66 | 13.61 | -6.66 |
| $F^- + CH_3Cl \rightarrow FCH_3 + Cl^-$ | -12.54 | -9.48 | 20.11 | -4.85 |
| $F^- \cdots CH_3Cl \rightarrow FCH_3 \cdots Cl^-$ | 2.89 | -4.57 | 29.62 | -4.42 |
| $OH^- + CH_3F \rightarrow HOCH_3 + F^-$ | -2.78 | -7.37 | 17.33 | -8.27 |
| $OH^- \cdots CH_3F \rightarrow HOCH_3 \cdots F^-$ | 10.96 | -5.72 | 47.20 | -3.26 |
| $H + N_2 \rightarrow HN_2$ | 14.69 | -10.52 | 10.72 | -0.98 |
| $H + CO \rightarrow HCO$ | 3.17 | -6.89 | 22.68 | 1.45 |
| $H + C_2H_4 \rightarrow CH_3CH_2$ | 1.72 | -6.29 | 41.75 | 1.28 |
| $CH_3 + C_2H_4 \rightarrow CH_3CH_2CH_2$ | 6.85 | -6.40 | 32.97 | -2.31 |
| $HCN \rightarrow HNC$ | 48.16 | -1.88 | 33.11 | -0.96 |

Table SVIII. SCAN errors for the S22 set of weak interactions [69]. For other functionals, see Ref [34]. The accurate reference values are from Ref. [69]. Unit: kcal/mol.

| **Systems** | Accurate | SCAN |
|---|---|---|
| NH3 dimer (C2h ) | 3.17 | -0.03 |
| H2 O dimer (Cs ) | 5.02 | 0.37 |
| Formic acid dimer | 18.8 | 1.83 |
| Formamide dimer (C2h ) | 16.12 | 0.27 |
| Uracil dimer (C2h ) | 20.69 | -0.36 |
| 2-pyridone–2-aminopyridine (C1 ) | 17 | -0.31 |
| Adenine–thymine WC (C1 )d | 16.74 | -0.86 |
| *CH4 dimer(D3d)* | 0.53 | -0.16 |
| *C2H4 dimer(2d)* | 1.5 | -0.43 |
| *Benzene-CH4(C3)* | 1.45 | -0.56 |
| *Benzene dimer(C2h)* | 2.62 | -1.48 |
| *Pyrazine dimer (Ch2)* | 4.2 | -1.49 |
| *Uracil dimer (C2)* | 9.74 | -1.74 |
| *Indole-Benzene(C1)* | 4.59 | -2.40 |
| *Adenine–thymine (C1 )* | 11.66 | -2.97 |
| C2H4-C2H2 | 1.51 | -0.16 |
| Benzene-H2O | 3.29 | 0.01 |
| Benzene-NH3 | 2.32 | -0.32 |
| Benzene-HCN | 4.55 | -0.47 |
| Benzene-dimer | 2.71 | -1.21 |
| Indole-Benzene(Cs) | 5.62 | -1.55 |
| Phenol dimer | 7.09 | -1.18 |

Table SIX. Errors in the equilibrium lattice constants (Å) of the LC20 solids [78] for SCAN and for three fully-nonlocal vdW-DF's (optPBE, optB88, and optB86b). For other functionals, see Ref [78]. The zero-point anharmonic expansion (ZPAE) was subtracted from the experimental zero-temperature values to yield the static-lattice "experimental" values (as in Ref. [27]). The data for the three vdW-DF's are from Ref. [50]. ME is the mean error, and MAE the mean absolute error.

| Solid | Expt. | optPBE | optB88 | optB86b | SCAN |
|---|---|---|---|---|---|
| Li | 3.451 | -0.011 | -0.019 | 0.001 | 0.009 |
| Na | 4.207 | -0.012 | -0.038 | -0.016 | -0.017 |
| Ca | 5.555 | -0.053 | -0.105 | -0.090 | -0.014 |
| Sr | 6.042 | -0.063 | -0.125 | -0.121 | 0.043 |
| Ba | 5.004 | -0.017 | -0.087 | -0.098 | 0.048 |
| Al | 4.019 | 0.039 | 0.035 | 0.017 | -0.014 |
| Cu | 3.595 | 0.060 | 0.037 | 0.010 | -0.028 |
| Rh | 3.793 | 0.050 | 0.038 | 0.012 | -0.004 |
| Pd | 3.876 | 0.084 | 0.065 | 0.033 | 0.019 |
| Ag | 4.063 | 0.111 | 0.078 | 0.038 | 0.016 |
| C | 3.555 | 0.030 | 0.022 | 0.017 | -0.005 |
| SiC | 4.348 | 0.038 | 0.027 | 0.021 | 0.001 |
| Si | 5.422 | 0.054 | 0.038 | 0.025 | 0.002 |
| Ge | 5.644 | 0.149 | 0.118 | 0.081 | 0.029 |
| GaAs | 5.641 | 0.142 | 0.110 | 0.076 | 0.017 |
| LiF | 3.974 | 0.093 | 0.059 | 0.063 | 0.006 |
| LiCl | 5.072 | 0.081 | 0.042 | 0.031 | 0.008 |
| NaF | 4.570 | 0.123 | 0.077 | 0.088 | 0.013 |
| NaCl | 5.565 | 0.108 | 0.057 | 0.062 | 0.002 |
| MgO | 4.188 | 0.064 | 0.043 | 0.042 | 0.018 |
| ME | | 0.054 | 0.024 | 0.015 | 0.007 |
| MAE | | 0.069 | 0.061 | 0.047 | 0.016 |